\documentclass[sigconf]{acmart}
\renewcommand\footnotetextcopyrightpermission[1]{}
\usepackage{booktabs} 
\usepackage{color}
\usepackage{listings}
\usepackage{graphicx}
\usepackage{hyperref}
\usepackage{url}
\usepackage{pgfplots}
\usepackage{xcolor}
\usepackage{xparse}
\usepackage{xspace}
\usepackage[htt]{hyphenat}
\usepackage{pifont}
\usepackage{algorithm}
\usepackage[noend]{algpseudocode}
\usepackage{balance}

\renewcommand{\textrightarrow}{$\rightarrow$}

\definecolor{codegreen}{rgb}{0,0.6,0}
\definecolor{codegray}{rgb}{0.5,0.5,0.5}
\definecolor{codepurple}{rgb}{0.58,0,0.82}
\definecolor{backcolour}{rgb}{0.95,0.95,0.92}

\lstdefinestyle{mystyle}{
    backgroundcolor=\color{backcolour},
    commentstyle=\color{codegreen},
    keywordstyle=\color{magenta},
    numberstyle=\tiny\color{codegray},
    stringstyle=\color{codepurple},
    basicstyle=\footnotesize,
    breakatwhitespace=false,
    breaklines=true,
    captionpos=b,
    keepspaces=true,
    numbers=left,
    numbersep=5pt,
    showspaces=false,
    showstringspaces=false,
    showtabs=false,
    tabsize=2
}

\colorlet{punct}{red!60!black}
\definecolor{background}{HTML}{EEEEEE}
\definecolor{delim}{RGB}{20,105,176}
\colorlet{numb}{magenta!60!black}

\lstdefinelanguage{json}{
    literate=
     *{:}{{{\color{punct}{:}}}}{1}
      {,}{{{\color{punct}{,}}}}{1}
      {\{}{{{\color{delim}{\{}}}}{1}
      {\}}{{{\color{delim}{\}}}}}{1}
      {[}{{{\color{delim}{[}}}}{1}
      {]}{{{\color{delim}{]}}}}{1},
}

\lstdefinelanguage{Ini}
{
    columns=fullflexible,
    morecomment=[s][\color{Orchid}\bfseries]{[}{]},
    morecomment=[l]{\#},
    morecomment=[l]{;},
    commentstyle=\color{gray}\ttfamily,
    morekeywords={},
    otherkeywords={=,:},
    keywordstyle={\color{green}\bfseries}
}
\newcommand{\ie}{i.e.,\@\xspace}
\newcommand{\eg}{e.g.,\@\xspace}
\newcommand{\etal}{et al.\@\xspace}

\newcommand{\etc}{etc.\@\xspace}
\newcommand{\gras}[1]{{\bf #1}}
\newcommand{\noindgras}[1]{\noindent{\bf #1}}

\newcommand{\projectname}{\texttt{CamQuery}\xspace}
\newcommand{\camflow}{\texttt{CamFlow}\xspace}

\renewcommand\footnotetextcopyrightpermission[1]{}

\begin{document}

\copyrightyear{2018}
\acmYear{2018}
\setcopyright{acmcopyright}
\acmConference[CCS '18]{2018 ACM SIGSAC Conference on Computer and Communications Security}{October 15--19, 2018}{Toronto, ON, Canada}
\acmBooktitle{2018 ACM SIGSAC Conference on Computer and Communications Security (CCS '18), October 15--19, 2018, Toronto, ON, Canada}
\acmPrice{15.00}
\acmDOI{10.1145/3243734.3243776}
\acmISBN{978-1-4503-5693-0/18/10}

\fancyhead{}

\title{Runtime Analysis of Whole-System Provenance}

\newlength{\alen}
\settowidth{\alen}{\it University of Cambridge}
\addtolength{\alen}{8mm}

\author{Thomas Pasquier}
\authornote{Part of this work was completed at Harvard University and at the University of Cambridge.}
\affiliation{
	\institution{University of Bristol}
}

\author{Xueyuan Han}
\affiliation{
	\institution{Harvard University}
}

\author{Thomas Moyer}
\affiliation{
	\institution{University of North Carolina at Charlotte}
}

\author{Adam Bates}
\affiliation{
	\institution{University of Illinois at Urbana-Champaign}
}

\author{Olivier Hermant}
\affiliation{
	\institution{MINES ParisTech\\ PSL Research University}
}

\author{David Eyers}
\affiliation{
	\institution{University of Otago}
}

\author{Jean Bacon}
\affiliation{
	\institution{University of Cambridge}
}

\author{Margo Seltzer}
\affiliation{
	\institution{University of\\ British Columbia}
}

\renewcommand{\shortauthors}{T. Pasquier \etal}

\begin{abstract}

Identifying the root cause and impact of a system intrusion remains
a foundational challenge in computer security.
{\it Digital provenance} provides a detailed history of the flow of information within a computing system, connecting suspicious events to
their root causes.
Although existing provenance-based auditing techniques provide
value in forensic analysis, they assume that such analysis takes
place only retrospectively.
Such post-hoc analysis is insufficient for realtime security
applications;
moreover, even for forensic tasks, prior provenance collection systems
exhibited poor performance and scalability, jeopardizing the
timeliness of query responses.

We present \projectname, which provides inline, realtime provenance
analysis, making it suitable for implementing security applications.
\projectname is a Linux Security Module that offers support for both userspace and in-kernel execution of analysis applications.
We demonstrate the applicability of \projectname to a variety of runtime security applications including 
data loss prevention, intrusion detection, and regulatory compliance.
In evaluation, we demonstrate that \projectname reduces the latency of realtime query mechanisms, while imposing minimal overheads on system execution.
\projectname thus enables the further deployment of provenance-based technologies to address central challenges in computer security.

 \end{abstract}
\maketitle

\section{Introduction}
\label{sec:introduction}
Timely investigation of system intrusions remains a notoriously difficult challenge~\cite{xu2009detecting, yuan2012improving, mace2015pivot}.
While security monitoring tools provide an initial notification of foul play ~\cite{b2000,gcl2008,sfl+2014,vvk+2004,yoo+2013,zcd+2012}, these indicators are rarely sufficient in and of themselves.
Instead, crafting an appropriate response to a security incident often requires scouring terabytes of audit logs to determine an adversary's method of entry, how their reach spread through the system, and their ultimate mission objective.
Such investigations not only require a human-in-the-loop, but are excruciatingly slow, at times requiring months of investigation and thousands of employee hours~\cite{k2010}.
This delay between an event's occurrence and its diagnosis represents a tremendous window of opportunity for attackers -- as they continue to exploit the system, defenders are still just getting their bearings.

{\it Digital provenance} (or \emph{provenance} for short) refers to the data being used in a variety of ways to address the challenges of forensic audits. By parsing individual records into causal relationship graphs that describe a system's execution, provenance enables defenders to leverage the full historical context of a system and to reason about the interrelationships between different events and objects. With provenance, forensic investigations can trace back a given security indicator (e.g., a port scan) to the attacker's point of entry (e.g., a malicious email attachment) \cite{king2003backtracking} and then trace forward from the entry point to determine what other actions the attacker has taken on the system.

Unfortunately, provenance-based auditing's growing popularity has uncovered significant limitations in its performance and scalability.
Early efforts to integrate provenance querying into production systems indicated that, even for modestly small organisations (e.g., 150 workstations), forensic queries can take on the order of hours or days to complete~\cite{priotracker}.
In an actual attack scenario, where a timely incident response could make the difference between victory and defeat, such delays are unacceptable.
Moreover, to date, provenance-aware systems have supported causal reasoning only as an after-the-fact forensic activity~\cite{ko2011system};
this is unfortunate, because provenance is also invaluable to a variety of runtime security tasks such as access control~\cite{park2012provenance, nguyen2013provenance}, integrity measurement~\cite{564676}, and regulatory compliance~\cite{am2009,bmv+2013,ma2014,pasquier-ubi-2017}.
To date, the design of low latency mechanisms for realtime provenance analysis has not been given adequate consideration in the literature.

The goal of this work is to bridge the gap between runtime security monitoring and post-hoc forensic analysis.
In support of this goal, we consider methods for the deep integration of provenance capture and analysis within the operating system.
We introduce \projectname, a framework that supports runtime analysis of provenance and thus enables its practical use for a variety of security applications.
\projectname pairs a runtime kernel-layer reference monitor -- expanding and modifying \camflow~\cite{pasquier-socc2017} -- with a novel query module mechanism that enables runtime provenance analysis and even mediation of system events.
\projectname modules present a familiar vertex-centric API, as popularised by modern graph processing systems such as GraphChi~\cite{kyrola2012graphchi} and GraphX~\cite{gonzalez2014graphx}.
In these vertex-centric platforms, full-graph analysis routines are expressed in terms of a small program that runs in parallel on every vertex (node) in the system.
The graph-structured nature of provenance data makes this model a good fit and permits use of a familiar API.
While these applications run directly over the live provenance stream, provenance can be simultaneously persisted to facilitate additional post-mortem and/or forensic analysis.

To demonstrate the generality of \projectname, we consider several exemplar query applications in \autoref{sec:example}.
1) a data loss prevention scheme~\cite{bates2015trustworthy} popular in provenance-security based community;
2) a provenance based intrusion detection scheme;
3) a mechanism to apply constraints on information flow;
and 4) a provenance signature scheme.
These case studies illustrate the rich space of design possibilities that are enabled through runtime provenance analysis. The source code for \projectname, along with associated applications and datasets, is available at \url{http://camflow.org}.

\vspace{0.2cm}

\noindent This paper makes the following contributions:\\
\noindgras{\projectname}: We present the design and implementation of  \projectname, an analysis framework over a live provenance stream.\\
\noindgras{Whole-system provenance modelling}: our work is the first to provide automated modelling of whole-system provenance through static analysis of the Linux kernel source code.\\
\noindgras{Exemplar Applications}: We demonstrate \projectname's efficacy in security-related applications, such as preventing loss of sensitive data or assuring log integrity.\\
\noindgras{Performance Evaluation}: We rigorously evaluate the performance of \projectname to demonstrate its effectiveness in realistic operating environments.\\
\noindgras{Availability}: We released an open-source implementation of \projectname. Based on the Linux Security Modules framework, \projectname is immediately deployable on millions of systems worldwide.

\section{Background}
\label{sec:background}
To provide context for the rest of the paper, we first introduce the concept of
whole-system provenance and then outline some shortcomings of
existing systems.

\subsection{Whole-System Provenance}

\begin{figure}[t]
	\centering
	\includegraphics[width=\columnwidth]{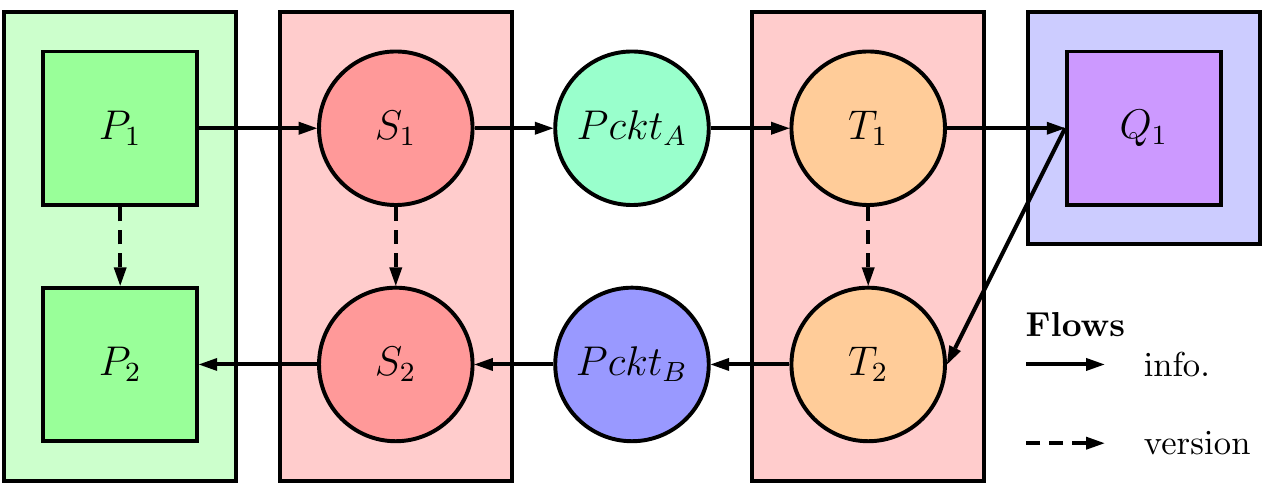}
	\caption{A simple provenance DAG: two processes ($P$ and $Q$) exchange packets ($Pckt_A$ and $Pckt_B$) through their respective sockets ($S$ and $T$).}
	\label{image:model:w3c}
\end{figure}

The W3C~\cite{moreau2013prov} defines provenance as a directed acyclic graph (DAG) where vertices represent \emph{entities} (data), \emph{activities} (transformations of data) and \emph{agents} (persons or organisations), and edges represent relationships between those elements.
\autoref{image:model:w3c} presents a simple example.
In our context, \emph{entities} are kernel objects, such as inodes, messages, and network packets;
\emph{activities} are tasks;
and \emph{agents} are users and groups.

In practice, it is impossible to represent a mutable process or file as a single vertex while simultaneously ensuring that the graph remains acyclic~\cite{Braun2006issues}.
For example, in a naive representation, a process that both reads and writes
a file immediately creates a cycle, because the process depends on the file
(due to the read), and the file depends on the process (due to the write).
Cycles are problematic.
Edges in the provenance graph represent dependencies between the states of different objects and express causal relationships. Therefore,
an object must depend only on the past (\ie the state of an object cannot depend on a future state).
The most commonly used cycle avoidance technique is to create multiple vertices per
entity or activity~\cite{muniswamy2006provenance}, each representing a version or state of the corresponding object.
We can see in \autoref{image:model:w3c} that new versions of the process $P$ and sockets $S$ and $T$ are created as information flows through these objects.

Using provenance graphs, we can detect and provide attribution for malicious behaviour~\cite{han2017} or actively prevent attacks using
provenance-based access control~\cite{nguyen2013provenance}.
However, using provenance to prevent actions requires that provenance is \emph{``complete and faithful to actual events''}~\cite{pohly2012hi}.
Missing events could sever connections, resulting in failure to reveal an important information flow; errant provenance could falsely implicate a benign process.
Pohly \etal~\cite{pohly2012hi} demonstrated that it was possible to satisfy such requirements by building provenance capture around the reference monitor concept~\cite{anderson1972computer} to mediate all events that should appear in the provenance graph.
They called this approach \emph{whole-system} provenance, which records events from system initialisation to shutdown.

\subsection{Issues With Provenance Architectures}
\label{sec:issue}
Existing provenance capture architectures were not designed with realtime support for security
applications in mind.
Therefore, unsurprisingly, they have
some fundamental limitations.
The traditional whole-system provenance capture stack, as first implemented in PASSv1~\cite{muniswamy2006provenance}, is built of the following five layers:
\begin{itemize}
	\itemsep0em 
	\item \gras{the capture layer} records system events;
	\item \gras{the collection layer} transports provenance information to where it may be used (\eg using messaging middleware such as Kafka~\cite{kafka} or Flume~\cite{flume});
	\item \gras{the storage layer} transforms system events into a provenance graph and persists it;
	\item \gras{the query layer} extracts provenance through queries relevant for a particular analysis;
	\item \gras{the analysis layer} interprets the provenance in the context of an application.
\end{itemize}

The use of whole-system provenance for runtime security applications is a relatively recent phenomenon.
Bates \etal~\cite{bates2015trustworthy} demonstrate provenance-based techniques to prevent loss of sensitive data in an enterprise, 
while Han \etal~\cite{han2017} use provenance to detect errant or malicious processes in a cloud environment.
Both of their systems were built on top of the conventional stack described above.
We argue that such an approach is suboptimal for provenance-based security applications,
incurring latency penalties arising from the need to store data before querying or analysing it.
Specifically, {\em when the goal of provenance analysis is mediation, 
delaying that analysis until after the data has been stored is impractical.}
Therefore, while existing architectures may be appropriate for post-mortem forensic investigations, they are not ideal for runtime security applications.
The goal of our work is to enable such applications through the introduction of vertex-centric, real-time analysis of streaming provenance.

\section{Runtime analysis framework}
\label{sec:query}
In the previous section, we made the case for realtime analysis over the provenance data stream.
We now present the design of \projectname, a framework for enabling such analysis to support runtime provenance-based security applications.

\begin{figure}[t]
	\centering
	\includegraphics[width=\columnwidth]{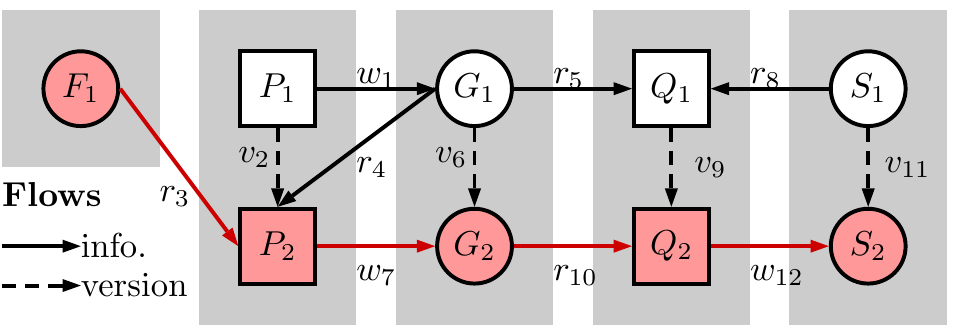}
	\caption{A demonstration of how path queries can be calculated through label propagation.
		The red (shaded) boxes indicate those vertices (with versions in subscripts) to which the ``confidential'' label is
		propagated.
		Confidential information flows from file $F$ to socket $S$, through process $P$, file $G$, and process $Q$. $r_i$, $v_i$, and $w_i$ stand for read, version, and write, respectively.
		The subscripts $i$ represent event ordering.}
	\label{image:implementation:graph}
\end{figure}

\subsection{Threat Model \& Assumptions}

We design \projectname with consideration for an adversary that has gained remote access to a host.
Once the adversary has gained access to the machine, they may engage in typical attacker behaviour such as installing malware, escalating their privilege level, or engaging in anti-forensic activities to hide evidence of their misdeeds.
However, we make the common assumption that the adversary does not have physical access to the machine.
Broadly, the goal of \projectname is to securely facilitate the provenance-based analysis of the adversary's actions in real time.

\noindgras{Trusted Computing Base (TCB):}
The TCB of \projectname includes a capture mechanism to generate a provenance graph from system events and a query mechanism to process the provenance at runtime, which we discuss at greater length in the remainder of this section.
Because any loaded kernel module is granted unrestricted access to kernel memory,
we assume that the entire kernel is distributed and installed in a trusted state,
which is a typical assumption in kernel-layer security mechanisms.
This assumption is made more reasonable through the use of integrity measurement techniques such as remote attestation and module signatures.
Protecting the capture mechanism from attackers who are able to alter kernel behaviour is an important but orthogonal issue that we discuss in \autoref{sec:challenges}.

\noindgras{Secure Provenance Store:}
If we wish to store provenance for post-mortem forensic analysis, an
adversary must not be able to corrupt it.
We assume the availability of secure provenance storage,
which can be achieved through a variety of known techniques.
For example,
Hasan~\etal~\cite{hasan09} present a hash-chain-based method for protecting provenance,
while Bates~\etal~\cite{bates2015trustworthy} secure provenance storage and transmission through the use of
type enforcement.
By layering these systems, it becomes possible to ensure full-stack trustworthy provenance.

\noindgras{Checkpointing}:
We assume that \projectname is deployed on a host that does not leverage checkpointing.
Checkpointing systems pose a challenge for \emph{all} provenance systems,
because restoring a checkpoint effectively moves a system backwards in time.
As a demonstrative example of this problem,
consider a policy to prevent conflicts of interest~\cite{brewer1989chinese},
\eg a policy to prevent a user who has read the Coca-Cola recipe from also reading the Pepsi recipe.
If checkpointing could be used to rollback the system to a state before the Coca-Cola recipe was read, an adversary could easily violate the policy.

\subsection{Motivating Example}
\label{sec:query:existing}

To identify the operational requirements of \projectname, we ground our discussion
in a prior example of provenance-based runtime security applications.
Bates~\etal~\cite{bates2015trustworthy} present a loss prevention scheme (LPS)
that disallows confidential information to be sent to an external IP address
by issuing provenance ancestry queries on all network transmissions.
Because this application was implemented on a conventional provenance capture stack,
query latency rapidly became the bottleneck --
even when the user queried a relatively small graph stored in an in-memory database,
the responses took upward of 21ms.
Worse, because response latency grew linearly with the size of the graph,
one would expect this application to quickly grow unusable under realistic conditions.

In contrast to ancestry queries, an alternative method of implementing LPS
would be to propagate security labels along the provenance graph in realtime,
as demonstrated in \autoref{image:implementation:graph}.
Because each object will be associated with the correct security label at the enforcement point,
graph traversal is no longer necessary and an authorization decision can be made in constant time.
Note that while this approach is akin to taint tracking,
a provenance-based approach allows for the expression of more complex queries than is possible in a conventional taint-tracking system.
With a provenance-based approach, we can express subtle propagation constraints based on properties of the graph (we demonstrate this in {\bf Example \#1} in \autoref{sec:example}). 
For example, Bates~\etal's LPS system propagates labels only along certain edge types, which is not possible in a data-centric taint-analysis system.

This approach to performing LPS can be generalized to a variety of other runtime security applications.\footnote{We return to the subject of example provenance-based security applications in \autoref{sec:example}.}
For example, in access control \cite{park2012provenance, nguyen2013provenance},
stream-based analysis can be used to express constraints on the properties of paths in the graph in a manner similar to computation tree logic
(\eg \emph{all} paths \emph{from} an external socket must not lead to disk \emph{until} they have gone through an anti-virus process)~\cite{emerson1982decision}.
Such constraints can be evaluated by building primitives above a value propagation algorithm.
This allows, for example, policies such as declassification
and path disjointedness to be built
to enforce conflict-of-interest constraints~\cite{brewer1989chinese}.
With this in mind, the goal of \projectname is to facilitate runtime security applications such as those considered above.
In addition, as we show in \autoref{sec:example}, the framework is sufficiently rich to be used, for example, to generate feature vectors in an intrusion detection setting.

\subsection{Overview}

\begin{figure}[t]
	\centering
	\includegraphics[width=\columnwidth]{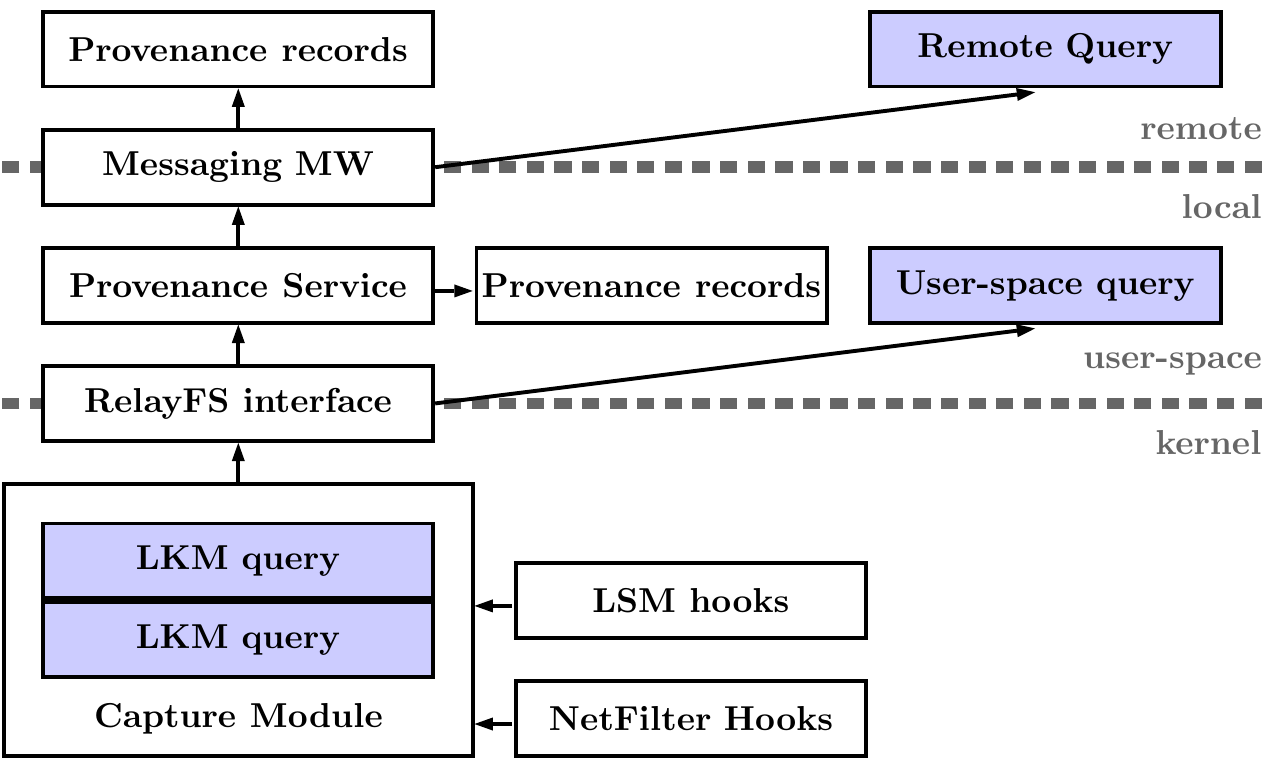}
	\caption{An overview of the \projectname framework.}
	\label{image:overview}
\end{figure}

\autoref{image:overview} presents an overview of the \projectname framework.
\projectname captures system events using \texttt{LSM} and \texttt{NetFilter} hooks; those events are transformed into a provenance graph within the \texttt{capture module} (\ie \camflow).
The \texttt{capture module} feeds graph elements (\ie edges and vertices) to stacked \projectname queries, which are either
built directly into the kernel or implemented as a loadable kernel module.
The kernel transfers these graph elements to user space for
1) consumption by user space queries; 2) recording for post-hoc analysis; or 3) transmission to a remote machine.
\projectname embodies the design goal of ensuring
a standard query implementation mechanism, independent of the three deployment options, discussed in \autoref{sec:implementation}.

\subsection{Provenance Monitor}

Like prior kernel-layer provenance capture systems (\eg LPM~\cite{bates2015trustworthy}, HiFi~\cite{pohly2012hi}), \camflow introduces a {\it provenance monitor} in the kernel.
A provenance monitor is a provenance capture mechanism that satisfies the reference monitor concept~\cite{formal:and1972}, possessing the properties of complete mediation, tamperproofness, and verifiability.
The relevance of these guarantees in the context of provenance capture is that they ensure that the provenance history is complete and accurate, even in the presence of an active attacker.
While past provenance monitors generally only denied system accesses if they were unable to generate a new record of the access (e.g., out of memory),
\projectname exposes a general mechanism for system mediation,
allowing security applications to authorize or deny new access based on the provenance history of the concerned principals.
Further details are in \autoref{sec:formalisation}.

\subsection{\projectname API}
\label{sec:implementation:algorithm}
\projectname provides an API, inspired by graph-processing frameworks such as GraphChi~\cite{kyrola2012graphchi} and GraphX~\cite{gonzalez2014graphx}, enabling straightforward implementation of value propagation applications. A query application consists of three functions:
\begin{enumerate}
	\itemsep0em 
	\item {\bf  init:} called upon query initiation to initialise the query's variables;
	\item {\bf out\_edge(v, e):} called on every outgoing edge $e$ of vertex $v$;
	\item {\bf in\_edge(e, v):} called on every incoming edge $e$ of vertex $v$.
\end{enumerate}

\projectname invokes
\texttt{out\_edge} and \texttt{in\_edge} in a manner
guaranteeing that edges are processed according to the partial order implied in paths in the graph and in topological order of the vertices.

\projectname calls the developer-defined
\texttt{out\_edge} and \texttt{in\_edge} functions with two parameters
containing \texttt{edge} and \texttt{node} data structures.
These structures expose attributes of the underlying kernel objects they represent (\eg inode, process, shared memory),
  allowing the developer to reference or modify the objects associated with the new system event. 
For example, the data structure representing a process vertex contains information such as UID, GID, namespaces, security context, system and user time, memory consumption, \etc;
  in turn, the edge data structure contains information such as offset, flags, mode, \etc.
There are around two dozen vertex types, e.g., path, network addresses, network packet, and shared memory states (complete list online~\cite{camflow_vertices}).
Similarly, there are over three dozen different edge types covering families of system calls (complete list online~\cite{camflow_relations}).
By specifying conditional constraints on the processing of vertex/edge labels and values,
  developers can express specific, complex queries.

\begin{lstlisting}[float=t,language=C, style=mystyle, caption={\projectname query in C.}, label=listing:implementation:query]
#define KERNEL_QUERY
#include "include/camquery.h"

static label_t confidential;

static void init( void ){
  confidential = get_label("confidential");
}

static int out_edge(union prov_msg* node, union prov_msg* edge){
	switch (edge_type(edge)){
		case RL_WRITE:
		case RL_READ:
		case RL_SND:
		case RL_RCV:
		case RL_VERSION:
		case RL_VERSION_PROCESS:
		case RL_CLONE:
			if ( has_label(node, confidential) )
				add_label(edge, confidential);
  }
  return 0;
}

static int in_edge(union prov_msg* edge, union prov_msg* node){
  if ( has_label(edge, confidential) ) {
    add_label(node, confidential);
    if( node_type(node) == ENT_INODE_SOCKET )
      return PROVENANCE_RAISE_WARNING;
  }
  return 0;
}

QUERY_NAME("My Example Query");
QUERY_DESCRIPTION("An example query");
QUERY_AUTHOR("John Doe");
QUERY_VERSION("0.1");
QUERY_LICSENSE("GPL");
register_query(init, in_edge, out_edge);
\end{lstlisting}

In addition to the manipulation of the provenance objects and existing kernel objects,
  \projectname also provides functions that allow developers to associate new labels or values with edges and vertices (\eg \texttt{add\_label}, \texttt{add\_ptr}).
\autoref{listing:implementation:query} shows a query that implements
  a loss-prevention scheme, which we describe at greater length in \autoref{sec:example}.
Associating labels with graph elements allows developers to easily implement, in a few lines of code, mechanisms such as taint tracking, information flow control, or access control.
Futhermore, using data structure association it is possible to build more complex graph analytics.
For example, we show in \autoref{sec:example} how to associate complex data structures with kernel objects and perform inlined computation while traversing the graph.
From that, we can compute, at runtime, feature vectors used to perform intrusion detection.

\projectname explicitly decouples the graph analysis implementation from the underlying kernel infrastructure.
The goal is to allow development of new provenance modules with a minimum of engineering effort.
For example, traditional taint tracking or information flow control implementations require extensive engineering effort~\cite{krohn2007information, roy2009laminar}, while it is possible to implement these applications in \projectname using only a few dozen lines of code.

\section{Implementation}
\label{sec:implementation}
We have implemented \projectname for Linux 4.14.15 and validated its use on Fedora 27.
The work presented here is fully implemented, used in multiple research projects, and is available online on GitHub (\url{https://github.com/CamFlow}) under a GPL v2 license.

\subsection{Capture Mechanism}
\label{sec:implementation:capture}

We built \projectname on top of the \camflow provenance capture system
\cite{pasquier2015camflow, pasquier-socc2017, camflow},
our actively-maintained provenance monitor built as a stackable
Linux Security Module (LSM) \cite{morris2002linux}.
Compared to other existing capture techniques~\cite{muniswamy2006provenance, gehani2012spade},
an LSM-based approach ensures that \camflow can observe and mediate all information flows between processes and kernel objects~\cite{edwards2002runtime, jaeger2004consistency, ganapathy2005automatic, georget2017verifying} (see \autoref{sec:formalisation} for further discussion).

Recording exact interactions between shared states (\eg mmap files, shmem, \etc) is challenging.
\camflow records those interactions by
conservatively assuming that
information always flows between processes and shared states.
We represent shared states as entities.
In the provenance graph, we add a relation from a process to the associated shared states when it receives information (\eg reading a file),
and a relation from the associated shared states to the process when it sends information (\eg writing a file).
We track shared memory by parsing through the memory data structure (\texttt{mm\_struct}) associated with each \texttt{task}.
Additionally, we extended \camflow to track provenance at the thread level rather than the process level.
Note that \camflow is the first whole-system provenance capture mechanism to do so.
Process memory is represented as a shared state between threads in the provenance graph.
We made these changes on top of the original design of \camflow to obtain more accurate provenance and consequently more accurate results in security applications such as intrusion backtracking~\cite{king2003backtracking}.
However, conservatively assuming the existence of information flows can lead to false positives.
We discuss this limitation and its potential solutions in \autoref{sec:challenges}.

To support runtime analysis, further changes to \camflow were necessary.
Existing provenance capture mechanisms, including past versions of \camflow,
do not directly generate graph elements in the kernel but instead generate logs of events that are processed in user space as part of the storage layer~\cite{muniswamy2006provenance, muniswamy2009layering,  gehani2012spade, pohly2012hi, bates2015trustworthy}.
We extended \camflow to generate the graph directly at the point of capture for two reasons:
1) event ordering is easier, as opposed to previous systems' complex computations to reconstruct kernel states and event orderings in user space;
2) more importantly, event ordering is made a precondition of the graph analysis in kernel space.

We modified the capture mechanism to embed limited provenance metadata alongside kernel objects to perform cycle avoidance in the kernel~\cite{muniswamy2006provenance, muniswamy2009layering}.
The cycle avoidance algorithm is entirely based on local properties of a node (\ie information about incoming and outgoing information flows) and does not require maintenance of any global state.
Fundamentally, we create a new version any time an object that sent information receives new information.
This guideline guarantees global acyclicity and avoids the creation of
a new state of an object that depends on the future.

Finally, we modified \camflow to publish graph components (\ie edges and vertices) as the system executes, while providing the following two partial ordering properties:
1) all incoming edges to a vertex are published before any outgoing ones;
2) edges and vertices along a path are published in order.
\projectname processes edges and vertices as they are published.

\subsection{Ensuring Completeness and Accuracy}
\label{sec:formalisation}

The design and implementation of \projectname extend the guarantees of past provenance monitors to support runtime provenance analysis.
The introduction of a query mechanism, which is described below, can
be used to further restrict system access.
The standard mechanisms used to secure the deployment of past provenance monitors are applicable to our system.
It naturally follows that \projectname possesses the same security properties as do past provenance monitors,
Therefore, we omit a complete security analysis,
and instead refer interested readers to the work of Bates \etal~\cite{bates2015trustworthy} for a detailed analysis of the security properties of provenance monitors.

Past provenance monitor implementations (\eg Hi-Fi~\cite{pohly2012hi} and LPM~\cite{bates2015trustworthy}) derive security properties from the guarantees provided by the formal verification of LSM placement~\cite{edwards2002runtime, zej2002, jaeger2004consistency, ganapathy2005automatic}, ensuring that they capture all interactions between kernel objects.
We extend this prior assessment of provenance completeness and accuracy:

\noindgras{Completeness: }We want to ensure that all flows of information between kernel objects are properly recorded.
The LSM framework~\cite{morris2002linux} was originally implemented to support Mandatory Access Control (MAC) schemes but not information flow tracking.
Recent work by Georget \etal~\cite{georget2017verifying, georget2017information} demonstrated, through static analysis of the kernel code base,
that the LSM framework is applicable to information flow tracking,
and that by adding a small number of LSM hooks, it was possible to properly intercept all information flows between kernel objects.
Building on their work, we maintain a patch~\cite{thomas_pasquier_2018_1319100} to the LSM framework
that allows {\camflow}, and by extension {\projectname}, to provide stronger guarantees than do previous whole-system provenance capture mechanisms.

\begin{figure}[t]
	\centering
	\includegraphics[width=0.7\columnwidth]{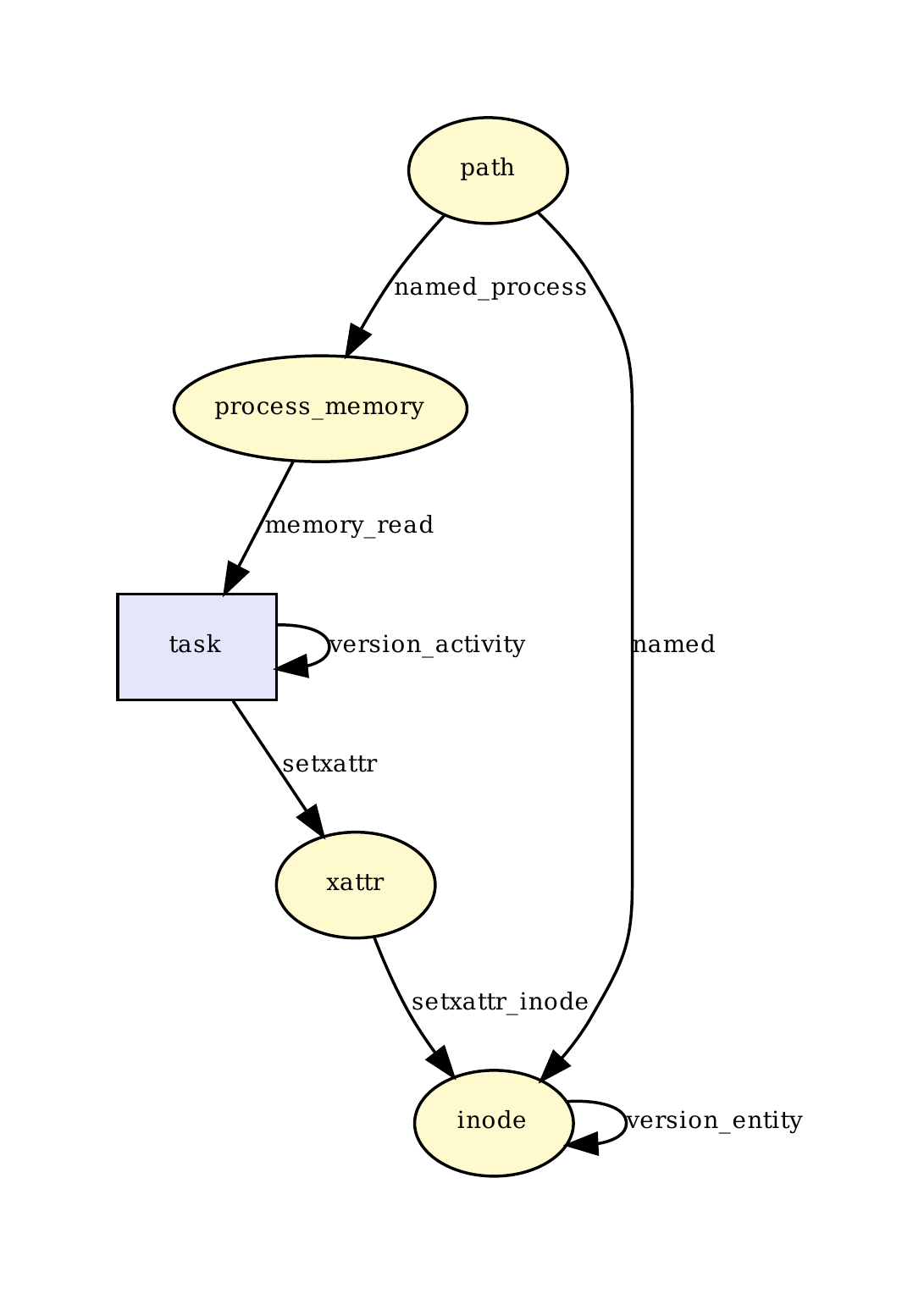}
	\caption{Provenance model for the \texttt{inode\_post\_setxattr} hook.}
	\label{image:inode_post_setxattr}
\end{figure}

\begin{figure}[t]
	\centering
	\includegraphics[width=\columnwidth]{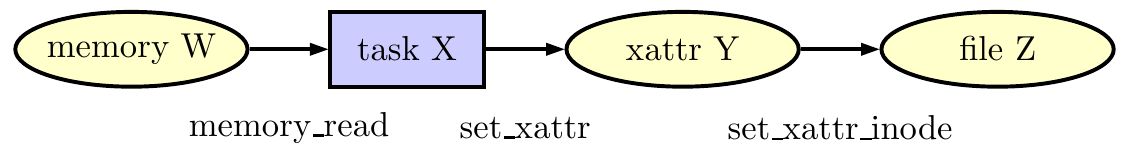}
	\caption{A whole-system provenance subgraph representing a valid instance of the model shown in \autoref{image:inode_post_setxattr}.}
	\label{image:instance}
\end{figure}

\noindgras{Accuracy: }We also provide accuracy guarantees for the recorded provenance.
We automatically analyse kernel source code to model the provenance generated by any \camflow-supported LSM hook (see \autoref{image:inode_post_setxattr} and \autoref{image:instance} for an example of such a model).
We then manually verify that all models meet our expectations.\footnote{Unfortunately, manual verification currently requires significant knowledge of the Linux kernel.}
Finally, through static analysis,
we identify the LSM hooks associated with each system call
and generate the associated provenance model, which we again manually verify.
This process is embedded in our continuous integration testing, with
results automatically updated in our Git repository~\cite{camflowreports} so that as the capture mechanism and the underlying kernel evolve,
we ensure the accuracy of our provenance capture.
We welcome meaningful scrutiny by third parties.
We believe this work is the first attempt towards formalisation of whole-system provenance.

We continue work on automated and formal analysis of whole-system provenance
capture.
Our future plans include combining static analysis techniques with dynamic provenance model benchmarking, as described in \eg Chan \etal~\cite{chan2017expressiveness}. 
Although we currently just assume a correct implementation of the ordering
properties described in \autoref{sec:implementation:capture}, our goal is to
formalise these as well.

\subsection{\projectname Query Configurations}
Depending on the security and performance requirements of a deployment, it may
not always be practical to embed analysis applications in the kernel.
For example, computationally expensive analysis may affect system stability, or
a proprietary analysis tool may need to be run on a separate host from the capture point.
Therefore, our implementation supports a variety of different configuration options that enable built-in kernel level analysis,
loadable kernel module analysis,
local user-level analysis, and
remote user-level analysis on a machine subscribed to the provenance stream.
While all of the deployment options run the same code,
only the in-kernel implementations can \emph{prevent} policy violations;
like previous systems, the user space and remote implementations
can only \emph{detect} violations after the fact.

\noindgras{Kernel-Based Configurations: }
\label{sec:implementation:kernel}
\projectname implements in-kernel queries using either Loadable Kernel Modules (LKMs) or directly linked objects.
LKMs are dynamically loaded object files that run in kernel space and have access to a subset of the kernel API.
Directly linked objects allow for shipping queries as part of the kernel.

Loading a query LKM invokes the \texttt{register\_query} function, which in turn invokes the \texttt{init} function.
After registration, the kernel invokes
\texttt{out\_edge} and \texttt{in\_edge} whenever \camflow records a new event.
Given the partial ordering property of our collection,
a vertex $v$ will receive all values propagated through an in-edge before
\texttt{out\_edge} runs on its outgoing edges.
If several queries are loaded, they execute sequentially in their load order.
These functions are actually executed before the actions they describe,
because they are executed from LSM framework hooks designed to implement Mandatory Access Control schemes.
This enables \projectname to prevent policy violations rather than merely detecting them.

\projectname maintains approximately 20 bytes of provenance state for kernel objects, \eg \texttt{inode}, \texttt{cred}.
By associating provenance with the kernel objects themselves, queries have access to the kernel objects, granting them visibility into kernel states.
Provenance for long-lived kernel states, such as \texttt{inodes}, persists across reboots through the use of extended attributes.

While the focus of this paper is enabling runtime query and analysis,
we observe that our framework creates opportunities at other layers of the provenance stack as well.
For example, we were able to rewrite \camflow's optional selective capture mechanism~\cite{pasquier-socc2017} using \projectname to reflect the modular nature of this component.
This mechanism makes it possible to limit provenance captured to a process,
an object, or characteristics of the provenance graph, \eg
recording the actions of only those processes belonging to a specific SELinux context e.g., to track the actions of an \texttt{httpd} server.

\noindgras{User space Configurations: }
\label{sec:implementation:stream}
The user space implementations of \projectname operate similarly to the kernel one.
Rather than producing an LKM, user-level queries produce a service that reads provenance records from either \texttt{relayfs} or a messaging middleware.
Queries process the stream by placing records into a sorted in-memory edge list and a persistent vertex map.

The \projectname capture mechanism writes records to per-core \texttt{relayfs} files
that are read in per-core batches,
producing a collection of partially-ordered edge lists that are not necessarily totally ordered.
To facilitate ordered processing of edges, a user space utility performs a merge of the per-core lists as follows --
for an out-edge $e$ received at time $t$,
all in-edges must have been received by $t+T$,
where $T$ is the QoS threshold.
At regular time intervals, the query processes all the edges satisfying $t<now-T$.

Rather than using timestamps to order edges, we use edge IDs;
the capture mechanism guarantees that edge ID ordering respects
the ordering properties described in \autoref{sec:implementation:capture}.
In a similar manner, we use provenance DAG causality relationships on network packets to produce a partial order across machines.
We then merge the per-core edge lists and the network packets to produce a sorted edge list.
The query processes each
edge sequentially by invoking the \texttt{in\_edge} and \texttt{out\_edge}
functions.

In addition to maintaining a list of edges, a user space query maintains a map of vertices.
We discard an edge after it is processed; we discard a vertex
either after processing an edge referencing a new version of
the vertex or after terminating events specific to the object (\eg a network packet will not be referenced after it has been received or a process will not be referenced after it has been terminated).
We show in \autoref{sec:evaluation} that, in practice, this represents a small memory footprint.

Vertex garbage collection relies on the semantics of system events.
We therefore record events relating to the life cycle of long-lived objects (\eg representing in the graph process kernel data structures being freed).
These events are not necessarily pertinent to the tracking of information flows, but greatly help with garbage collection.
If the framework were to be applied to other types of provenance (\eg Spark provenance~\cite{interlandi2015titian}), the garbage collection algorithm would require different domain knowledge.

Converting the code in \autoref{listing:implementation:query} to
its user space equivalent is trivial.
We modify Line 1 to reflect the proper target, currently one of
\texttt{MW\_QUERY} or \texttt{RELAY\_QUERY}, indicating from
where the service will obtain data (a middleware-provided data stream or \texttt{relayfs}, respectively).
We add two more query attributes after line 39.
\texttt{QUERY\_MSG} specifies the messaging middleware broker address and topic.
Note that although the kernel transmits information to \texttt{relayfs} before executing the
action corresponding to the query, as we do not control when user processes are scheduled, we
cannot guarantee that the query service has an opportunity to process the provenance before
the corresponding action is taken.
As such, the user level implementation, and by extension the distributed one, can guarantee only violation detection, not prevention.

\noindgras{Discussion: }
The different guarantees available from different \projectname configurations provide a rich set of trade-offs.
While in-kernel queries have access to the underlying kernel data structures and can prevent events from occurring, they incur overhead on every system call.
\autoref{sec:evaluation} illustrates this power/performance trade-off.
On the other hand, user space queries can perform runtime monitoring only, raising alerts relatively quickly, but not quickly enough to prevent events from occurring.
However, such queries can build on existing libraries to \eg perform log analysis~\cite{pasquier-socc2017}.
Additionally, the scheduler is responsible for scheduling user space queries, so it can more easily adjust to system workload as shown in \autoref{sec:evaluation}.

\section{Example Applications}
\label{sec:example}
We designed \projectname to enable development of important security and compliance applications, such as intrusion detection~\cite{han2017, han2018provenance}, enforcement of software licenses, and compliance with data regulation~\cite{pasquierinformation, pasquier-ubi-2017}.
During development, we implemented several algorithms inspired by the literature to validate the suitability of the framework.
In this way, we ensured that we could implement meaningful provenance analysis at runtime in widely different use cases.
We provide the examples below to illustrate the range of applications that can be implemented with \projectname.

\begin{figure}[t!]
	\centering
	\includegraphics[width=0.7\columnwidth]{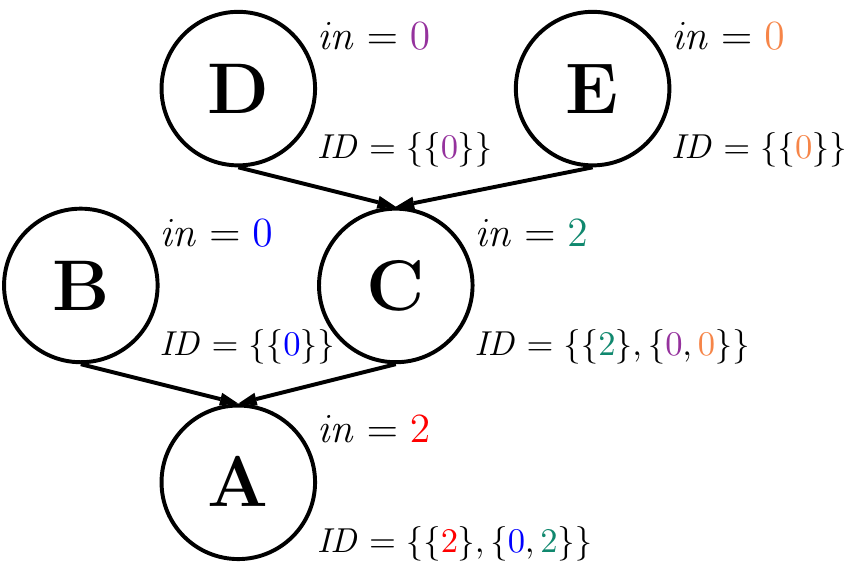}
	\caption{Calculating vertices' structural identity (Depth=1).}
	\label{image:idvec}
\end{figure}

\begin{algorithm}[t!]
	\caption{Encoding Structure Identity (pseudo code).} \label{alg:struct}
	\begin{algorithmic}[1]
		\Function{out\_edge}{vertex, edge}
		\State Calculate DTW between its own $\mathit{ID}$ and parent  $\mathit{ID}$
		\State Publish feature vector
		\State Write to the edge its own $\mathit{ID}$
		\EndFunction

		\Function{in\_edge}{edge, vertex}
		\State Increment vertex in-degree counter
		\State Read and save parent $\mathit{ID}$
		\State Merge parent $\mathit{ID}$ to build own $\mathit{ID}$
		\EndFunction
	\end{algorithmic}
\end{algorithm}

\noindgras{Example \#1: Data Loss Prevention.}
\label{sec:implementation:code}
We first demonstrate how the framework works with a relatively simple graph processing algorithm implementing the loss prevention scheme of Bates \etal~\cite{bates2015trustworthy},
which prevents sensitive data from leaving a system (as discussed in \autoref{sec:query:existing} and shown in \autoref{image:implementation:graph}).

\autoref{listing:implementation:query} shows the implementation of this application.
The query contains four main components: the \texttt{init} function (lines 6--8),
the \texttt{out\_edge} function (lines 10--23),
the \texttt{in\_edge} function (lines 25--32),
and a set of query attribute statements (lines 34--39).
Before query registration, \projectname executes the query attribute statements to set the query's properties exactly once.
Then, during registration, it calls the \texttt{init} function exactly once.
Subsequently, \projectname invokes
\texttt{in\_edge} and \texttt{out\_edge} for every edge in every active query.

The LPS scheme considers only certain flows of information as meaningful in the context of the policy.
Therefore,
  it propagates labels (lines 19--20) only over the relevant flows (line 11--18),
  raising a warning if the label ever reaches a socket (lines 26--29).
In more complex scenarios,
developers can maintain global states within a query or associate more complex data structures with edges or vertices.
Notably, we emphasise that \autoref{listing:implementation:query} contains all of the
  required runtime logic for a label-based loss-prevention system, demonstrating the efficiency with which
  security applications can be expressed in \projectname.
Outside of this application, our LPS scheme assumes only:
1) a labelling state that tags sensitive information sources with the \emph{confidential} label,
2) that correctness requires handling explicit information flow only, not side channels, and
3) that sensitive information that reaches a system exit point (\eg a socket) raises a warning.

We can design more complex algorithms around programmable label propagation.
An example extension uses label propagation to detect abnormal behaviour in a system.
For example, one can easily use \projectname to track the origin of executables and sensitive data as previously labelled.
An indicative abnormal system behaviour might be an executable that did not originate from a trusted repository manipulating sensitive data.
Once a potentially harmful pattern is detected,
techniques such as intrusion backtracking~\cite{king2003backtracking} can be used to manually assess the situation.
Other more sophisticated, automated techniques are also available;
we refer interested readers to the work by Eshete~\etal~\cite{eshete2016attack}, which describes, in more depth, use cases of provenance for label-based intrusion detection techniques.

\noindgras{Example \#2: Intrusion Detection.}
Recent work explores how to improve the efficacy of Intrusion Detection Systems (IDS) using provenance~\cite{han2017}.
With this work as inspiration, we show
how to implement anomaly detection using \projectname.
Provenance-based intrusion detection is still a nascent field
that has not yet been demonstrated to be robust against a realistic active adversary;
we use it merely as a demonstration of {\projectname}'s
ability to allow for the construction of complex feature vectors.

Our proposed approach to provenance-based intrusion detection is based on unsupervised learning techniques.
Our goal is to learn how the system normally behaves, build a model of such behaviour, and detect large deviations from the model.
We generate provenance graphs from the executions of our system in a controlled environment under normal conditions.
As in previous work~\cite{han2017, han2018provenance}, we capture provenance data during multiple runs of a cloud application under a representative workload
and build a model of normal behaviour.

Our example IDS uses a replicator neural network~\cite{hawkins2002outlier} (RNN, also known as an autoencoder) to detect anomalies in a graph.
An RNN consists of an encoder and a decoder.
The encoder performs compression of the feature vector.
The decoder then reconstructs the input feature vector from the compressed vector.
The objective of training is to minimise the distance between the input of the encoder and the output of the decoder.
RNNs are often used for outlier
detection, as they often have difficulty reconstructing feature vectors that diverge significantly from the training dataset.
In our case, we leverage this behavior to detect abnormal structures in the provenance graph.
Using {\projectname}, we construct a feature vector for every vertex,
which is composed of the following three parts:
1) vertex attributes (\eg vertex type, security context, UID, namespace, \etc);
2) changes of some attributes over time (\eg UID, memory or CPU usage for processes, \etc);
and 3) the structural identity~\cite{ribeiro2017struc2vec} of the vertex, which represents the graph structure surrounding the vertex.

Structural identity is a vectorisation of the graph neighborhood,
which represents the context in which a vertex exists,
and is critical for anomaly identification in outlier detection~\cite{ienco2017semisupervised} and intrusion detection~\cite{chandola2009anomaly}.
We define a neighborhood as the $n-\mathit{ancestry}$ of a vertex,
because descendants are unknown when we generate feature vectors at runtime.
The structural identity is built from ancestor in-degrees.
For each vertex, we maintain a list, $L$, of length $n+1$.
Let $i$ be the 0-based index of each element of this list.
$L_0$ is the in-degree of the vertex itself and
$L_i, i > 0$ is an in-degree sequence, a list consisting of the in-degrees of
the $i\mathrm{th}$ generation ancestors.
Thus, a vertex with two parents, each of which has no ancestors, is associated with the following list: $\{\{2\},\{0,0\}\}$;
\autoref{image:idvec} shows a concrete example.

Following Ribeiro \etal~\cite{ribeiro2017struc2vec},
we use Dynamic Time Warping (DTW), a technique for calculating the similarity between two temporal sequences~\cite{berndt1994using}, to calculate the distance between two \emph{in-degree sequences}.
We then populate the feature vector of a vertex with each of the DTW distances between a vertex and its ancestry.
This set of distance is the structural identity of the vertex.

\projectname propagates in-degree sequences along each path of a provenance graph.
Using the \texttt{out\_edge} function, each vertex passes its in-degree sequence to its descendants.
A child vertex receives sequences from all of its parent vertices and updates its own sequences using the \texttt{in\_edge} function.
Algorithm~\autoref{alg:struct} illustrates this.
When the \texttt{out\_edge} function runs, the vertex contains enough information to calculate its structural identity.

\begin{table}[t!]
	\centering
	\resizebox{\columnwidth}{!}{
		\begin{tabular}{l|cc}
			Vulnerability ID & Detection rate  & False positive  \\
			\hline
			\texttt{MariaDB} race condition exploit~\cite{CVE-2016-6663} & 100\%   & 0\%  \\
			\texttt{MySQL} root privilege escalation~\cite{CVE-2016-6664} & 50\%  & 0\%  \\
			\texttt{Nagios} core root privilege escalation~\cite{CVE-2016-9566} & 90\%   & 0\%  \\
		\end{tabular}
	}
	\caption{Preliminary results for our \projectname IDS mecahnism.}
	\label{table:ids}
\end{table}

\autoref{table:ids} shows some preliminary results of the intrusion detection scheme.
We generate training data by executing unexploited instances of each vulnerable application.
We then test the IDS by running a collection of normal and abnormal application executions.
While a full-fledged evaluation of our IDS mechanism is beyond the scope of this paper,
  we measure the computational cost of feature vector generation in \autoref{sec:evaluation}.

\begin{figure}[t]
\begin{flushleft}
$\mathrm{Q}1$) \textbf{path existence}\\
$\exists p: A \Rightarrow B$;\\
$\mathrm{Q}2$) \textbf{existence of a vertex on all paths between two vertices}\\
$\forall p: A \Rightarrow B, \exists v \in p, v \neq A $ AND $v \neq B$;\\
$\mathrm{Q}3$) \textbf{absence of a vertex on all paths between two vertices}\\
$\forall p: A \Rightarrow B, v \notin p$;\\
$\mathrm{Q}4$) \textbf{path disjointedness}\\
$\forall v \in p, v \notin p'$;\\
$\mathrm{Q}5$) \textbf{constraints on properties and types in a path}\\
$\forall v \in p$, if $v_\mathit{type}$ = T then $P$($v$),
for a specified property P and type T.
\end{flushleft}
\caption{\projectname can be used to query a variety of information flow properties. Here, we denote
a path from vertex $A$ to vertex $B$ as $p: A \Rightarrow B$.}
\label{fig:if-queries}
\end{figure}

\noindgras{Example \#3: Information Flow.}
\projectname can execute single-pass algorithms that rely on value propagation along paths in the graph.
For example, we implemented the simple primitives summarised in Figure \ref{fig:if-queries}.
Each implementation required just a few dozen lines of C code.
The data loss prevention scheme introduced in Example {\#}1, for example, tests for path existence ($\mathrm{Q}1$).

Using these queries, \projectname can aid in the enforcment or auditing of regulatory compliance.
The Sarbanes-Oxley act (SOX) applies to publicly held US corporations. The intent of the law is to establish security controls and accountability of personnel to protect against data tampering to hide fraud. While the law itself does not specifically address computing systems, every major corporation today relies heavily on computers to process financial data and report to the Securities and Exchange Commission (SEC). Specifically, Sections 302 and 404 detail the required safeguards for data to ensure accuracy in financial reporting and required disclosures. To be SOX compliant, an organization must carefully consider and have policies for data creation, publishing, retention, access, distribution, and lifecycle.

We consider here just three of the cases mentioned above. The first control is data access (Section 302.4.B), which requires that companies have controls in place to track accesses to data and ensure that company officers are aware of all relevant data.
The provenance records kept as forensic evidence ensure full compliance with the requirement to track data access.
A report detailing all the data entities appearing in the captured provenance could inform company officers of the ``relevant'' data.
Additionally, corporations could instantiate policies to detect accesses that do not comply with the act.

The second control we consider is data creation and the ability for a reporting officer to attest that the reported information is valid. This requires that data must not be tampered with before reports are created and filed with the SEC.
We can write \projectname policies that restrict data access to only those users and activities involved in report generation.
There are multiple ways to express this,
  one of which would be to label activities that are known to be acceptable, then write policies that verify that all activities between data generation and the SEC filing are labelled as such.
This is a query of type $\mathrm{Q}5$.
An alternative is to label all unacceptable techniques, \eg using a text editor on the data, and check that no such activities appear in the path between the data and SEC filing.
This is a $\mathrm{Q}3$ type of query.

Sarbanes-Oxley Title V deals with analyst conflicts of interest. It requires financial analysts to disclose conflicts of interest, ensuring that investors are not being misled by the biases of a financial analyst. These conflicts of interest can be avoided using separation of concerns policies~\cite{brewer1989chinese} that create information barriers preventing the exchange of information that would produce a conflict of interest. As a specific example, consider a financial analyst who is working with one company (Company A) as part of a hostile takeover of another company (Company B). Information concerning the takeover must not be transmitted to the brokerage department that could use the information to alter customer investments to increase profits for the financial company.
This is a $\mathrm{Q}4$ type of query.

\noindgras{Example \#4: Graph Integrity.}
To ensure the integrity of our provenance graph,
we implemented the directed acyclic graph signature scheme proposed
by Aldeco-P{\'e}rez \etal~\cite{aldeco2010securing}.
This technique is often cited in the literature as a solution to provenance integrity.

\begin{figure}[t]
	\centering
	\includegraphics[width=\columnwidth]{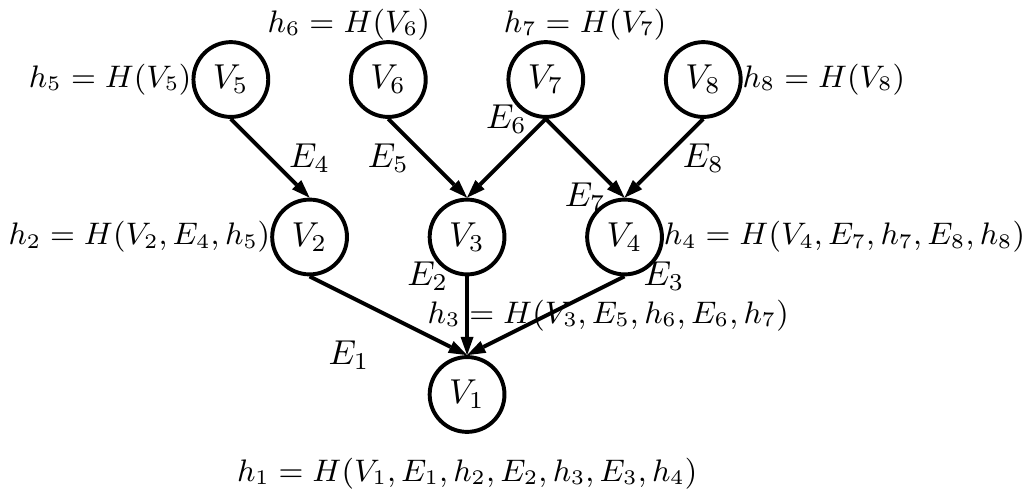}
	\caption{\projectname can be used to assure integrity by generating a signed provenance graph.}
	\label{image:sign}
\end{figure}

The system generates a chain of hashes according to the graph structure, as shown in \autoref{image:sign}.
The capture mechanism then signs these hashes.
The analysis engine can re-calculate the hashes for a graph to verify that they correspond to the signed value.
An advantage of this scheme is efficient verification, as it is not necessary to verify the entire graph to verify vertex $V$.

We leverage a kernel keyring infrastructure for key management (we took inspiration from eCryptfs~\cite{halcrow2005ecryptfs}) and the cryptographic API to perform related operations.
The resulting solution is a heavyweight, in-kernel query in the evaluation in \autoref{sec:evaluation}.
While it was easy to implement graph signing in \projectname,
unsurprisingly, creating signatures on every system call incurs significant overhead, even when the cryptographic algorithm itself is relatively lightweight.
Our measurements suggest that the provenance graph signature scheme~\cite{aldeco2010securing} is impractical at scale and inadequate when whole-system provenance capture is considered.
It also serves as a cautionary tale: while it is easy to implement a variety of applications using \projectname, not all such applications will exhibit acceptable performance.
Creating provenance integrity schemes that are practical at scale is an important open problem beyond the scope of this paper.

\section{Experimental Evaluation}
\label{sec:evaluation}
\begin{figure*}[t]
\includegraphics[width=\textwidth]{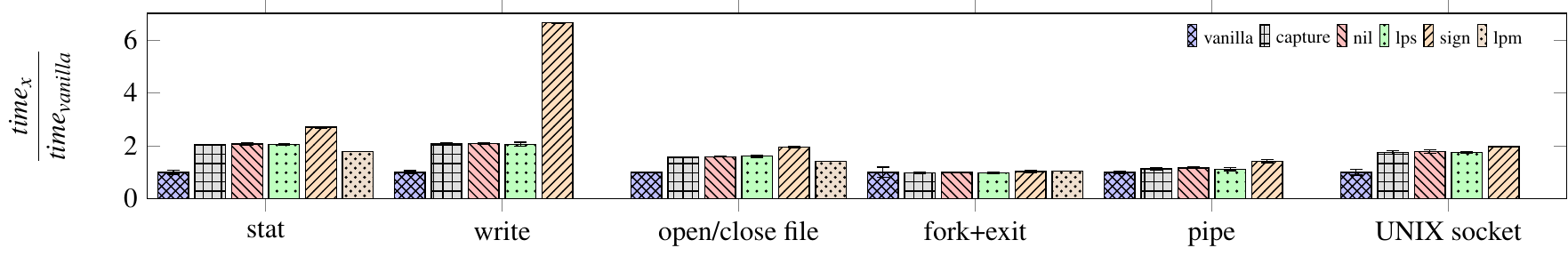}
\caption{Normalised overhead of queries (LPM capture overhead as reported in~\cite{bates2015trustworthy} is given when available).}
\label{figure:evaluation:perf}
\end{figure*}

We evaluate both the in-kernel and local user-space implementations to determine
how much overhead \projectname introduces and how that overhead is split between
provenance capture and query support.

We use workloads derived from those found in the whole-system provenance literature to provide meaningful points of comparison.
We run the benchmarks on a bare metal Fedora 27 machine, with Linux kernel 4.14.15 and \camflow 0.3.10 with an Intel i7-7700 2.8 GHz CPU and 32 GiB of RAM.
Due to space constraints, we present only a subset of our results.
Instructions on obtaining our code and reproducing all our results are available online (\url{http://camflow.org}) following recommendations by Collberg \etal~\cite{collberg2016repeatability}. 
Throughout the evaluation, we refer to the following setup:

\noindgras{vanilla: }unmodified Linux 4.14.15 kernel;

\noindgras{capture: }whole-system provenance capture;

\noindgras{nil: }nil in-kernel query (\texttt{in\_edge} and \texttt{out\_edge} simply return zero);

\noindgras{lps: }the loss prevention scheme in-kernel query described in~\autoref{sec:example} Example {\#}1;

\noindgras{sign: }the provenance signature in-kernel query described in~\autoref{sec:example} Example {\#}4;

\noindgras{ids: }the user-space query building feature vectors for the IDS described in~\autoref{sec:example} Example {\#}2.

\subsection{In-kernel Queries}
\label{sec:evaluation:performance}

\begin{table}[t]
\begin{tabular}{ l | c| c | c | c | c}
  Test Type & \gras{vanilla} & \gras{capture} & \gras{nil} & \gras{lps} & \gras{sign} \\
  \hline
  \multicolumn{6}{c}{Process tests, times in $\mu s$, smaller is better}\\
  \hline
  stat 							& 1.20 	& 2.44 	& 2.48 & 2.46 	& 3.24\\
	read 							& 0.22 	& 0.35 	& 0.35 	& 0.36 	& 1.05\\
	write 						& 0.15	& 0.31 	& 0.32 	& 0.31 	& 1.01\\
  open/close file 	& 2.04 	& 3.21 	& 3.24 	& 3.28 	& 4.00\\
  fork+exit					& 87.6 	& 85.5 	& 86.6 	& 85.7	& 89.8\\
  fork+shell				& 862 	& 860 	& 866 	& 855   & 861\\
  \hline
  \multicolumn{6}{c}{Latencies in $\mu s$, smaller is better}\\
  \hline
  pipe							& 3.47 	& 3.92	& 4.05	& 3.88 & 4.91\\
  UNIX socket				& 3.70 	& 6.44 	& 6.61 	& 6.47 & 7.28\\
\end{tabular}
\caption{LMbench measurements.}
\label{table:eval:lmbench}
\end{table}

\noindgras{Micro-benchmark: }
We used LMbench~\cite{mcvoy1996lmbench} to illustrate the impact of the provenance capture and query on system call performance.
\autoref{table:eval:lmbench} and
\autoref{figure:evaluation:perf}
present a subset of LMbench results.
Our provenance capture overhead is comparable to that reported for other systems~\cite{pohly2012hi, bates2015trustworthy}.
This is as expected and provides a sanity check.
More interesting is that the addition of online querying introduces relatively little overhead.

Indeed, execution time of a single system call is equal to
$V_s + n_s(C + Q) + m_s C$, where
$V_s$ is the execution time of the system call $s$ on a vanilla kernel.
$n_s$ is the number of edges in the graph corresponding to the system call $s$ (\eg a socket \texttt{send event} contains at least 2 edges, one from the process to the socket, and the other from the socket to the packet, and potentially edges corresponding to kernel object versions).
$m_s$ is the number of vertices in the graph corresponding to the system call $s$.
$C$ is the cost of capture and
$Q$ is the cost of the query.
The relative overhead is higher when $V_s$ is small, as C and Q are independent of the underlying system call execution time.
The overhead of LPM~\cite{bates2015trustworthy} (and of other previous provenance capture systems \eg~\cite{pohly2012hi, muniswamy2006provenance}) is $V_s + C_s$\ where $C_s$ is the cost of capturing the event corresponding to $s$, as LPM records system events rather than directly producing the graph structure (see \autoref{sec:implementation:capture}).

One of the advantages \projectname provides over prior work is a
drastic reduction in the time between an attack and its detection.
Bates~\etal~\cite{bates2015trustworthy} reported that it took their system 21ms to evaluate the same policy and further noted that ``these results are highly dependent on the size of the graph. [Their] test graph, while large [6.5 million vertices, and 6.8 million edges], would inevitably be dwarfed by the size of the provenance on long-lived systems''~\cite{bates2015trustworthy}.
The authors suggested that the performance could be further improved by using deduplication~\cite{xie2013evaluation} and pre-pruning techniques~\cite{bates2015take, pasquier-socc2017}.
However, they did not evaluate the performance impact of such improvements.
They did, however, report that graph size can be reduced by up to 89\% through pre-pruning techniques~\cite{bates2015take, bates2017taming}.
Even if we assume that the reduction produces a proportional improvement in query time,
the resulting 2.31ms per query is several orders of magnitude larger than the overhead imposed by \projectname for a similar application (\texttt{lps} in \autoref{table:eval:lmbench}).

\begin{figure}[t]
	\centering
	\includegraphics[width=\columnwidth]{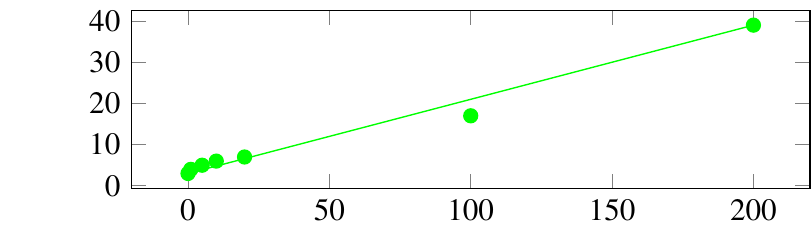}
	\includegraphics[width=\columnwidth]{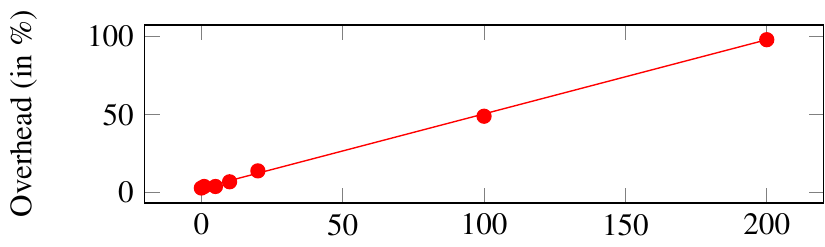}
	\includegraphics[width=\columnwidth]{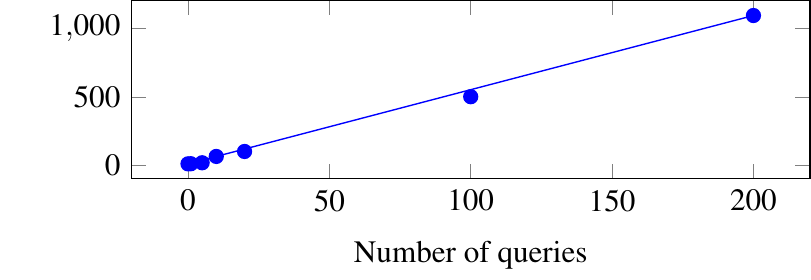}
	\caption{Benchmark results (\texttt{unpack} in green/top, \texttt{build} in red/middle and \texttt{postmark} in blue/bottom) as a function of the number of active queries (we run 0 to 200 concurrent \texttt{lps} queries). Note the difference in the $y$ axes for the different benchmarks.}
	\label{image:evaluation:queries}
\end{figure} 
\begin{table*}[t!]
\begin{tabular}{ l | c | c | c | c |  c || c c}
  Test Type & \bf vanilla & \bf capture & \bf nil & \bf  lps & \bf  sign & \bf PASS & \bf LPM\\
  \hline
  \multicolumn{8}{c}{Execution time in seconds, smaller is better}\\
  \hline
  unpack 				& 14.98	& 15.48 (3\%) & 15.63 (4\%) & 15.76 (5\%) & 16.68 (11\%) & NA & NA\\
  build 				& 402	& 411 (2\%) & 416 (3\%) & 417 (3\%) & 448 (11\%) & 15.6\% & 2.7\%\\
  \hline
  \multicolumn{8}{c}{4kB to 1MB file, 10 subdirectories,}\\
  \multicolumn{8}{c}{4k5 simultaneous transactions, 1M5 transactions}\\
  \hline
  postmark 				& 127 	& 145 (14\%) & 144 (13\%) & 146 (15\%) & 226 (78\%) & 11.5\% & 7.5\%\\
  \hline
\end{tabular}
\caption{Macro-benchmark results. PASS~\cite{muniswamy2009layering} and LPM~\cite{bates2015trustworthy} overhead as reported by the authors.}
\label{table:eval:macrobenchmark}
\end{table*}

\noindgras{Macro-benchmark:}
We contextualise the significance of the overhead measured in the micro-benchmarks using the Phoronix test suite~\cite{phoronix}.
We select benchmarks commonly used in the system provenance literature.
Consistent with the micro-benchmark results, the
macro-benchmark results (\autoref{table:eval:macrobenchmark}) show that provenance capture introduces negligible overhead for the kernel build benchmark and up to 15\% overhead for Postmark.
For reference, we also include reported overheads for prior systems (PASS and LPM).
As the Linux kernel versions
(2.6.x for the two mentioned systems vs 4.14.15 for \projectname)
and the underlying hardware vary greatly across these evaluations,
the results simply provide context and suggest that \projectname exhibits capture overhead comparable to prior systems.
The overhead is higher for benchmarks where the number of system calls per unit of time is larger, as the overhead is only incurred on interactions between a process and the system call interface.

\noindgras{Query stacking:}
The prior results show that a single query introduces acceptable overhead; next
we assess the impact of an increasing number of queries
executing concurrently.
We run the macro-benchmarks from \autoref{table:eval:macrobenchmark} with a varying
number of active queries and show the results in
\autoref{image:evaluation:queries}.
On the positive side, overhead increases linearly with the number of queries.
On the negative side, the Postmark overhead is particularly high, because it is a system-intensive workload, and system calls
trigger query evaluation.
While \texttt{build} and \texttt{unpack} spend approximately 10\% and 18\% of their time, respectively, in the kernel,
Postmark spends 85\% of its time in the kernel, making 253,000 system calls per second (over twice the rate of the other benchmarks).
It should be noted that production systems running hundreds of queries is unrealistic.
Further, we plan to explore the possibility of merging a set of queries into a single module, with the goal of reducing the number of redundant operations. This is a non-trivial task left for future work.

\begin{figure}[t]
	\centering
	\includegraphics[width=\columnwidth]{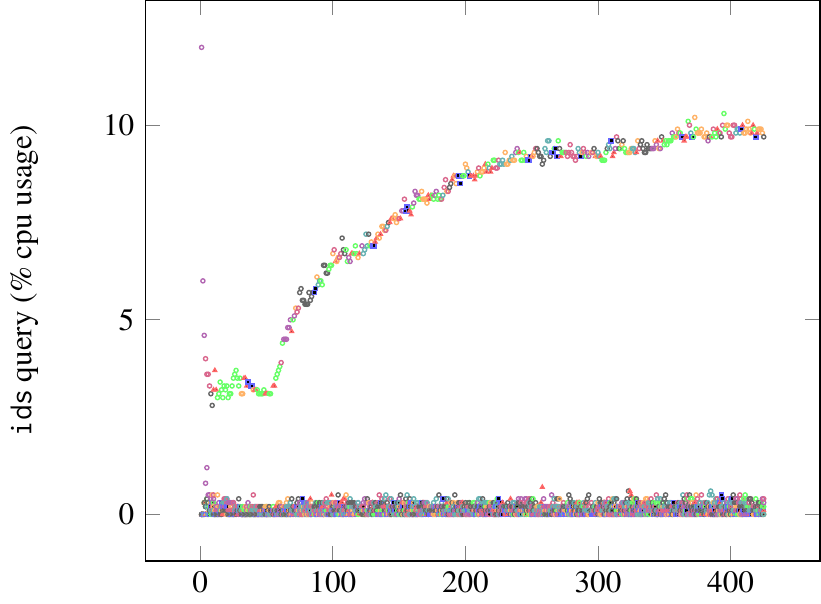}
	\includegraphics[width=\columnwidth]{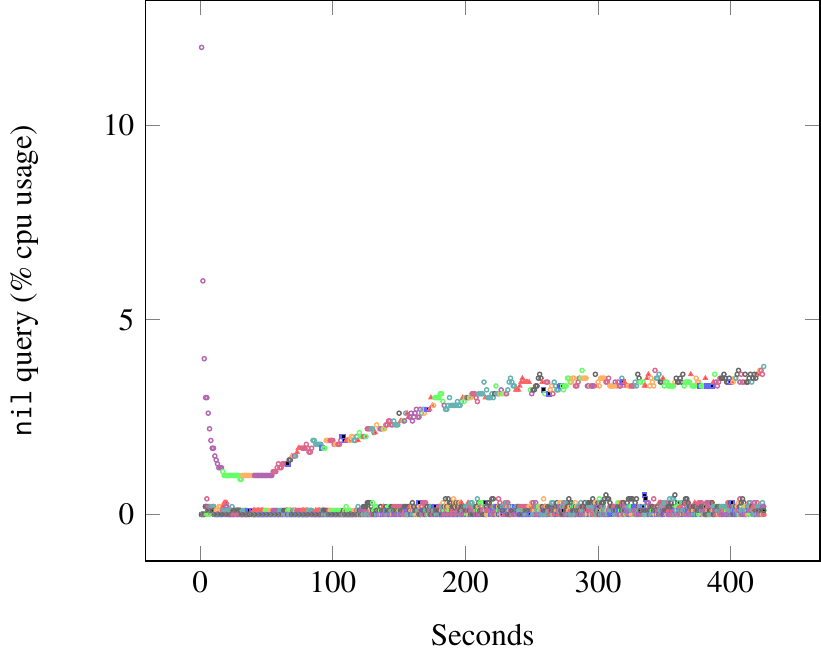}
	\caption{Percentage of CPU usage per core (each colour represents one of the eight cores) used by the \texttt{ids} (top) and \texttt{nil} (bottom) queries during a kernel compilation.}
	\label{image:evaluation:cpu}
\end{figure}
 
\subsection{User space queries}
\label{sec:evaluation:query}

\begin{table}[h]
	\resizebox{\columnwidth}{!}{
		\begin{tabular}{ l | c | c c | c c}
			Test Type & vanilla & \bf in-kernel & overhead & \bf userspace & overhead \\
			& & & (over capture) & & (over capture) \\
			\hline
			\multicolumn{6}{c}{Execution time in seconds, smaller is better}\\
			\hline
			unpack 				& 14.98	& 15.76 & \bf 5\% & 15.91 & \bf 6\% \\
			build 				& 402	& 417	& \bf 4\% & 427 & \bf 6\% \\
			\hline
			\multicolumn{6}{c}{4kB to 1MB file, 10 subdirectories,}\\
			\multicolumn{6}{c}{4k5 simultaneous transactions, 1M5 transactions}\\
			\hline
			postmark 				& 127 	& 146 	& \bf 15\% & 147 & \bf 15\%\\
			\hline
		\end{tabular}
	}
	\caption{Overhead of the \texttt{lps} query when compiling the Linux kernel.}
	\label{table:eval:macrobenchmark2}
\end{table}

Next, we want to evaluate the performance impact of running queries in user space.
We compare the overhead of the \texttt{vanilla} and \texttt{lps} in-kernel configurations from the previous section to that of the \texttt{lps} user space configuration, where the query is run as a \texttt{systemd}
managed service running on the same machine as the workload, reading provenance from \texttt{relayfs}.
\autoref{table:eval:macrobenchmark2} shows the results for the Linux kernel
\texttt{unpack} and \texttt{build} benchmark and Postmark.
Note that the user space overhead is only minimally larger than the in-kernel
overhead.

We next investigate how user space queries impact system workload
by running the \texttt{nil} query and the complex \texttt{ids} query.
The \texttt{ids} query generates feature vectors used by a machine learning algorithm to perform intrusion detection.
We run the kernel \texttt{build} benchmark as our system workload, as it generates a relatively large and complex graph (just over 25 million edges were processed by each query) when compared with the other two benchmarks.
At regular intervals, we record the memory and per-core CPU consumption of the two queries.

\begin{figure}[t]
	\centering
	\includegraphics[width=\columnwidth]{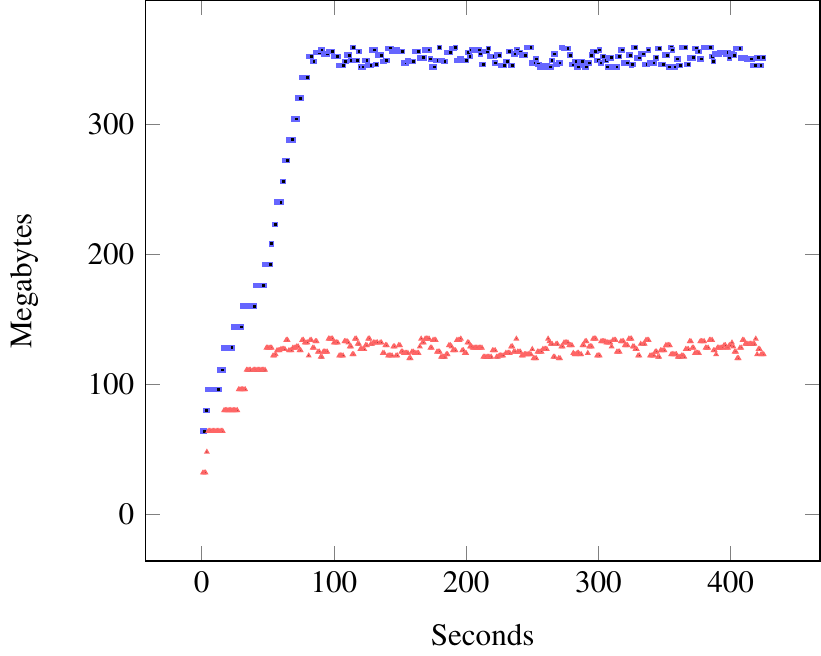}
	\caption{Memory used by the \texttt{ids} (blue/square) and \texttt{nil} (red/triangle) queries during a kernel compilation.}
	\label{image:evaluation:memory}
\end{figure}
 
In contrast to kernel queries, the user space query runs at regular intervals, processing all the newly arrived edges.
\texttt{Relayfs} creates a ring buffer mapped to a pseudofile per CPU core to transmit data to the query service.
The service runs one reader thread per core, reading the data from its \texttt{relayfs} file and populating the edge list and the vertex map.
Another thread, the processing thread, sorts the edge list and performs the query at regular time intervals.
As shown in \autoref{table:eval:macrobenchmark2},
the core running the processing thread reaches about 9\% utilisation for the \texttt{ids} query and 4\% for the \texttt{nil} query, while the
other cores, which are running reader threads, have a CPU utilisation between 0\% and 2\%.
The user space query competes with other workloads on the system for CPU time,
which may degrade application performance.
The multi-coloured nature of the lines in \autoref{table:eval:macrobenchmark2} shows that the processing thread moves among the cores.

\autoref{image:evaluation:memory} illustrates the memory consumption for the same queries.
The memory used by the user space query corresponds to the list of edges and vertices (which includes the propagated values).
The memory usage stabilises, as vertices are garbage collected, to around 305 MB for the \texttt{ids} query and 125 MB for the \texttt{nil} query.

\section{Challenges \& discussion}
\label{sec:challenges}
\projectname has limitations and raises interesting questions that go beyond the particular framework presented here.

\noindgras{Query Language: }
\projectname uses a programmable graph processing framework to express policies, rather than the seemingly more user-friendly DSL approach.
A DSL would undoubtedly need to be designed with a particular application in mind (\eg compliance enforcement, access control, \etc) and it would be challenging to make it amenable to queries such as the intrusion detection feature vector computation.
We believe that such languages are important but are part of individual applications rather than a general framework.
We plan to explore the design and development of a DSL for the provenance-based access-control scenario.
We note also that, concurrent to this study, Gao~\etal introduced SAQL~\cite{gxl+2018} and AIQL~\cite{gao2018aiql} which both introduce a domain-specific query language to aid in forensic investigation. 
They are implemented over data streams and persistent storages respectively.
We plan to explore how a similar language could be expressed through our vertex-centric query API.

\noindgras{Distributed Systems: } A challenge for \projectname is the ability to reason about computations that occur in a distributed system.
The user space implementation can be extended to support these systems with relative ease, but doing so eliminates the possibility of performing policy enforcement (see~\autoref{sec:implementation:stream}).
Supporting enforcement in a distributed system requires that the query be partitioned into per-machine segments combined into a kernel enforcement mechanism.
This partitioning necessitates the ability for the system to validate that the other machines in the system will accurately enforce the policy, \ie they are high-integrity and have the necessary enforcement mechanisms and provenance policies loaded.
Once a machine validates the integrity and suitability of a system, it must generate a ``proof'' that the policy has been enforced.

The ability to perform policy enforcement would open up new opportunities for \projectname in distributed settings, but also new challenges.
For example, it enables \projectname to function as the building block for a secure distributed taint propagation system
with the potential to allow sophisticated logic using complex labels.
To implement such a system, however, two important considerations must be taken into account, among others.
First,
we must ensure that \projectname is: minimally invasive, fully integrated into the existing network stack;
and is compatible with non-provenance-aware hosts,
especially if we hope to insert arbitrarily complex taint information in network packets.
Second,
transmission must be authenticated and tamperproof to \eg man-in-the-middle attack.
The latter might be addressed by existing secure network protocols such as IPSec, but technical challenges remain.

\noindgras{Trust: } The ability for a system to prove statements about its integrity and processing state is best suited to trusted computing, \eg trusted hardware and remote attestation.
In the above distributed system setting, there is a need for systems to generate ``proofs'' of their current state.
These proofs need to account for several system characteristics, including 1) the current integrity state of the system (hardware, firmware, software, \etc); 2) the currently loaded policies and; 3) the current state of the data being processed.
To prove the current integrity state of the system and the currently loaded policies, we can turn to techniques such as the Linux Integrity Measurement Architecture (IMA)~\cite{sailer2004design}.
IMA measures the load-time integrity of user space applications and files read by root.
These measurements are stored in the Trusted Platform Module (TPM) to support remote attestation, \ie generating an unforgeable proof of the measurements stored in the TPM.
The TPM is an inexpensive trusted hardware component that provides a small amount of protected storage for measurements and cryptographic keys.
These measurements can be signed by a key loaded into the TPM to support remote attestation, proving the current integrity state of the loaded system.
IMA will measure the policies being loaded as an LKM as long as the policy loading is done by root since the default policy measures all files read by root.
The remote attestation allows a remote verifier to determine the current state of the kernel and user-space applications.
What is still needed are mechanisms that enable a remote verifier to validate that the currently loaded policies are correctly enforced.

\noindgras{Storage: }The issue of storing provenance is orthogonal to the topic of this paper.
However, we believe that the work presented here represents a paradigm shift in provenance systems.
Whole-system provenance implementations have been faced with the issue of building a back-end that can
ingest high throughput~\cite{moyer2016high}, provide integrity and non-repudiability~\cite{Balakrishnan2017}, and handle large volumes of data while providing low latency queries.
Decoupling query performance from storage overhead introduces myriad new architectures for such systems.

\noindgras{False positives from flow tracking:} A well-understood limitation of the proposed approach is the potential for false positives when information flows are inferred.
For example, if a \texttt{task} reads from a file and writes to another, whole-system provenance capture systems conservatively assume that information was transferred,
even though it is not necessarily always the case.
Conservatively inferring information flow via shared memory is another major source of false positives.
Similar issues also arise in most system-level information flow control or taint tracking systems.
A potential solution to reduce the number of false inferences is to capture information flow within applications, using techniques such as bitcode transformation~\cite{tariq2012towards}, binary rewriting~\cite{cheng2006tainttrace, lzx2013-beep}, or static analysis~\cite{myers1999jflow}.
Such techniques are related to provenance layering~\cite{muniswamy2009layering}, the capture of internal application provenance alongside system level provenance to improve the accuracy of provenance records.
While the capture of such provenance is a well-understood problem, its analysis and scalability remain relatively unexplored.
 
\section{Related work}
\label{sec:rw}
We place this work both in the context of prior work on whole-system provenance capture and more general information flow tracking approaches, as
techniques such as Information Flow Control and Taint Tracking share many characteristics with provenance collection systems.

\noindgras{Provenance Systems.}
There have been several provenance capture mechanisms implemented in the Linux kernel~\cite{muniswamy2006provenance, pohly2012hi, bates2015trustworthy,mzx2016}.
LPM~\cite{bates2015trustworthy} uses provenance DAGs to enforce information flow constraints by querying graphs at sink points (\eg at the network interface).
The authors verify that paths from a source A to a sink B respect some well-defined properties expressed in the query.
However, their approach requires performing database queries where
query latency is a function of the graph size, which increases linearly over time.
Therefore, it suffers from a lack of scalability, slowing down over time as provenance accumulates.
\projectname addresses this issue by executing queries at runtime over the provenance stream, introducing bounded overhead independent from the graph size as shown in \autoref{sec:evaluation}.

\noindgras{Provenance Reduction.}
Recently, the issue of provenance storage and query performance has received considerable attention in the literature.
LogGC performs garbage collection on redundant events that have no forensic value~\cite{lzx2013},
while BEEP~\cite{lzx2013-beep} and MPI~\cite{mzw+2017} improve post-mortem analysis by solving the problem of dependency explosion.
PrioTracker~\cite{priotracker} accelerates forensic queries by prioritising the traversal of rare events in large provenance graphs.
These systems primarily exist at either the storage and query layer of the provenance stack;
in tackling the issue of log reduction through taint tracking, 
ProTracer employs a similar approach to \projectname by merging the capture and storage layers~\cite{mzx2016}.
While this work has led to dramatic improvement in the efficiency of provenance, \projectname achieves the orthogonal but interrelated goal of improving provenance performance through deep integration of analysis routines with the underlying capture framework.
An interesting avenue for future research would be considering how the above reduction techniques could be incorporated into the flattened provenance stack that \projectname envisions.

\noindgras{Provenance Applications.}
Provenance has been leveraged in the service of a variety of security applications.
Because provenance can be used to generate a model of known good executions of a system,
recent work has considered using provenance data to perform anomaly detection~\cite{han2017, hassan2018towards}.
Han \etal~\cite{han2017} use machine learning (ML) algorithms to detect outlier graph structures.
Hassan~\etal~\cite{hassan2018towards} use a graph grammar to build deterministic finite state automata and verify that the graph can be parsed.
Unlike the example shown in \autoref{sec:example}, where we generate data as the graph is being produced, they accumulate provenance to generate ``windows'' that are then analysed.
We have shown in~\autoref{sec:example} that it was possible to generate feature vectors for the ML-based approach.
It should also be possible to implement the graph parsing stage (\ie detection stage) of Bates~\etal's work using the \projectname framework.

Park~\etal~\cite{park2012provenance} formalise the notion of provenance-based access control (PBAC) systems along three dimensions:
1) the type of data used to make decisions (observed vs.~disclosed provenance~\cite{Braun2006issues});
2) object dependencies (information flow between objects) vs.~user dependencies (information flow between users); and
3) whether policies are available to the system or learnt through the traversal of provenance graphs;
\projectname, like most PBAC systems in the literature (~\cite{bates2013towards,sze2015provenance}), 
uses observed provenance, although it could be augmented by disclosed provenance.
Layering of provenance systems~\cite{muniswamy2009layering} could enable such a capability, although we are not aware of any layered PBAC enforcement model.
We plan to explore this approach in future work, with both application level~\cite{muniswamy2009layering} and network level provenance~\cite{zhou2011secure}.

\noindgras{Information Flow Control Systems.}
Previous work on information flow control enforcement at the OS level, such as HiStar~\cite{zeldovich2006making}, Flume~\cite{krohn2007information}, and Weir~\cite{nadkarni2016practical}, uses labels to define security and integrity contexts that constrain information flows between kernel objects.
Labels map to kernel objects, and a process requires decentralised management capabilities to modify its labels.
Point-to-point access control decisions are made to evaluate the validity of an information flow. Through transitivity, it is possible to express constraints on a workflow (\eg collected user information can only be shared with third parties as an aggregate).
SELinux~\cite{smalley2001implementing} provides a similar information flow control mechanism but without decentralised management.
A typical way of representing and thinking about information flow in a system is through a directed graph.
However, current object labelling abstractions do not take advantage of this representation, and it is difficult to reason about when defining policies.
\projectname differs from these systems in that it allows
the implementation of such mechanisms directly on the graph abstraction.

\noindgras{Taint Tracking Systems.}
Techniques such as ``colouring''~\cite{hwang2010trusted} or tainting~\cite{muthukumaran2015flowwatcher} of data and resources have been proposed as a means to detect data misuse.
TaintDroid~\cite{enck2014taintdroid} implements such an approach in the Android OS to detect applications disclosing personal information
to an unexpected third party (\eg disclosing the owner's contact list to advertisers).
\projectname can be used to achieve similar results as taint tracking systems but provides more control through its expressive query mechanism on how taints are propagated within the system.
Furthermore, the provenance records, kept as forensic evidence, provide a rich resource that can be mined to identify, understand, and explain the source of a disclosure.

\noindgras{Security Monitoring.}
In today's enterprise environments, security incidents occur when a primary indicator of compromise is triggered from security monitoring software such as an anti-virus detection alert or a blacklisted URL in the organisation's network logs~\cite{lyle2010trusted}.
In current security products, such indicators report only limited context as to the circumstances under which the alert occurred, \eg process ID or packet header information, but do not report the historical chain of events that led to the suspicious activity.
Past work has attempted to compensate for this lack of lineage through the fusion~\cite{b2000,gcl2008} or correlation~\cite{sfl+2014,vvk+2004,yoo+2013,zcd+2012} of multiple indicators of the compromise. However, it does not address the fundamental limitation that security monitoring tools lack
the ability to reason over the entire context of a system execution.
Thus, attack reconstruction has typically been relegated to (offline) forensic analysis~\cite{king2003backtracking,mlk+2015,lzx2013,mzw+2017,mzx2016,pfv2002,xwd+06,zzl+2014,zrl+2016}.
In contrast, \projectname provides a mechanism to build runtime security monitoring based on the entire history of system execution, thus representing a significant step forward compared to the state-of-the-art.

\section{Conclusion}
\label{sec:conclusion}
More than a decade ago, PASS~\cite{muniswamy2006provenance} represented a paradigm shift in how we think about provenance capture, moving from application-specific capture, to a system-wide holistic mechanism.
In this paper, \projectname rethinks how we envision provenance applications, severing the always-present, implicit link to database back-ends.
We make the distinction between runtime detection applications which should be built above live streams of provenance data to identify policy violations or anomalies, and forensic applications that run post-mortem, leveraging database support to provide explanations.
By drastically rethinking the conventional provenance architecture, we are able to reduce the time between an event (such as an attack, data leakage, non-compliance with regulation, \etc) and its detection, by several orders of magnitude, while simultaneously storing the data for post-mortem forensic investigation.
We continue to actively develop \camflow and \projectname as we investigate provenance applications.
The work is entirely open-source and we invite others to build upon it.

\section*{Availability}
\label{sec:availability}
The work presented in this paper is open-source and available for download at \url{http://camflow.org/} under a GPL v2 license.

\section*{Acknowledgments}
\label{sec:acknowledgments}
\noindent This work was supported by the US National Science Foundation under grants SSI-1450277 End-to-End Provenance, CNS-1750024 CAREER and CNS-1657534 Transparent Capture and Aggregation of Secure Data Provenance for Smart Devices.
Early versions of CamFlow open source software were supported by
UK Engineering and Physical Sciences Research Council grant EP/K011510 CloudSafetyNet.
 
\bibliographystyle{ACM-Reference-Format}
\balance
\bibliography{biblio,bates-bib-master}
\balance
\end{document}